\documentclass[12pt]{article}
\usepackage[T2A]{fontenc}
\usepackage[koi8-r]{inputenc}
\usepackage[english,russian]{babel}
\begin{document}
\large
\renewcommand
{\abstractname}{\bf Abstract}
\baselineskip 24pt
\begin {center} Yu.~P.~Chuburin
\end {center}
\begin {center}
{\bf
INVERSE SCATTERING PROBLEM FOR THE SCHR\" ODINGER
OPERATOR WITH A SEPARABLE POTENTIAL 
 }
\end {center}
\begin {abstract}
We prove that the inverse scattering problem for the Schr\"odinger operator with 
the separable potential  
can be reduced to the solving of a certain singular integral equation. We establish 
the uniqueness of the potential corresponding to given scattering amplitude
in the class of separable potentials.

\end{abstract}
\markright {}
\section {Introduction}

In this paper, we consider the Schr\"odinger operator in $L^2({\bf R}^3)$ of the 
form $H=H_0+V$, where $H_0=-\Delta , \, V=\lambda (\cdot ,\psi _0)\psi _0$ is a
separable potential with $\lambda \in {\bf R}$ and $ \psi _0(x)\in L^2({\bf R}^3)$. 
We assume that $V\neq 0$.

In what follows, we use the notation
$$
R_0(p)\phi (x)=\frac {1}{4\pi}\int _{{\bf R}^3}\frac {e^{ip|x-y|}}{|x-y|}
\phi (y)dy,        \eqno (1)
$$
where  $\phi \in L^2({\bf R}^3)\cap L^1({\bf R}^3), \, {\rm Im} \, p\geq 0.$
In the case $E=p^2\in {\bf C}\setminus [0,\infty )$ the operator of the
form (1) is the resolvent of the operator $H_0$ at the point $E$.

The Lippmann-Schwinger equation in the case under consideration has the form 
$$
\psi (x,k)=e^{ik\cdot x}-R_0(|k|+i0)V\psi (x,k)=
$$
$$
=e^{ik\cdot x}-\frac {\lambda (\psi ,\psi _0)}{4\pi}\int _{{\bf R}^3}
\frac {e^{i|k||x-y|}}{|x-y|}\psi _0(y)dy        \eqno (2)
$$
where $k\in {\bf R}^3$ and
 $$(\psi ,\psi _0)=\int _{{\bf R}^3}\psi (x)
\overline {\psi _0(x)}dx.$$ 
We consider this equation in the class $\psi \in L^{\infty}({\bf
R}^3)$.

We denote the scattering amplitude by
$$
f(|k|,n,\omega )=-\frac {\lambda (\psi ,\psi _0)}{4\pi}\int _{{\bf R}^3}
e^{-i|k|n\cdot y}\psi _0(y)dy        \eqno (3)
$$
where $n=x/|x|, \, \omega =k/|k|$  (see [1]).
By an inverse scattering problem we mean a finding of the potential
$V$ using the scattering amplitude $f$ given for all
(perhaps, sufficiently large) energies $E=k^2$. (This is  the third kind of the inverse 
problem in [1]).

In the sequel, we use the notation
$$
\hat {\phi}(k)=(2\pi )^{-3/2}\int _{{\bf R}^3}e^{-ik\cdot x}\phi (x)dx, \, 
\phi \in L^2({\bf R}^3)
$$
for the Fourier transform and $\check {\phi}$ for the inverse Fourier transform.
We shall denote by
$$
S\phi (x)=\frac {1}{\pi i}\int _{{\bf R}}\frac {\phi (y)}{y-x}dy
$$
the singular integral operator (here we regard the principal value 
of the integral). This operator acting, in particular, in the space 
$L^2({\bf R}^1)$ (see [2]). 
\section {Main equation}

We assume that the function $\psi _0$ satisfies the following conditions:
$$
(1+x^2)\psi _0\in L^2({\bf R}^3),\, \psi _0\in L^1({\bf R}^3)\cap L^{4/3}({\bf R}^3),
\eqno (4)
$$
$$
|{\rm grad}\, \psi _0|\in L^1({\bf R}^3)\cap L^{4/3}({\bf R}^3),\, \int _{{\bf R}^3}\frac 
{\psi _0(x)\overline {\psi _0(y)}}{|x-y|}dxdy\neq -\frac {4\pi}{\lambda}.
$$
Because of the first condition, there are exist the first continous
derivations of the function $\hat {\psi}_0$ in all variables. Making use of the second and 
the third conditions and also 
the Hausdorff-Young inequality [3], we obtain the inclusions
$$
\hat {\psi}_0(k)\in L^4({\bf R}^3)         \eqno (5)
$$
and
$$
|\widehat {\partial \psi _0/\partial x_j}(p)|=|p_j\hat {\psi}_0(p)|\in L^4({\bf R}^3), 
\, j=1,2,3.                                             \eqno (6)
$$

We need the following theorem (see [1]).
\vskip 0.5cm

{\sc Theorem 1.} {\it Assume that the relation
$$
1+\lambda (R_0(|k|)\psi _0,\psi _0)\neq 0.                      \eqno (7)
$$
holds. Then there exists the unique solution of the Lippmann-Schwinger equation (2) and}
$$
f(|k|,n,\omega )=-\frac {2\pi ^2\lambda \hat {\psi}_0(|k|n)\overline {\hat {\psi}_0(|k|\omega )}}
{1+\lambda (R_0(|k|)\psi _0,\psi _0)}.			\eqno (8)
$$
\vskip 0.5cm

We introduce the notation
$$
F(|k|)=\int _0^{2\pi}d\phi \int _0^{\pi}d\theta {\rm sin} \theta f(|k|,\omega ,\omega )
$$
where $|k|, \phi , \theta $ are the spherical coordinates of the vector $\omega$. Further, set
$$
\xi (|k|)=\pm |k|^2 \int _0^{2\pi}d\phi \int _0^{\pi}d\theta {\rm sin} \theta 
|\sqrt {\lambda }\hat {\psi}_0(|k|,\phi ,\theta )|^2.    \eqno (9)
$$
Here $\pm$ is the sign of $\lambda$ (the argument of the function $\hat {\psi}_0$ is 
written in the spherical coordinates). 
We assume that the function $\xi$ is extended on the real axis as even function.
\vskip 0.5cm		      

{\sc Lemma 1}{\it The following relation
$$
F(|k|)=-\frac {4\pi ^2\xi (|k|)}{|k|(2|k|+i\pi (1+S)(\xi )(|k|)}  \eqno (10)
$$
holds.}
\vskip 0.5cm

{\sc Proof}. It is obviously that the function $\xi (q)$ is continously differentiable
and summable on the real axis. Therefore, in the folowing transformations the Sohotskii-Plemelj
formula can be used . We also change the integration order and use that the
function $\xi $ is even. Thus, we have
$$
\lambda (R_0(|k|)\psi _0,\psi _0)=\lambda (R_0(|k|+i0)\psi _0,\psi _0)= \eqno (11)
$$
$$
=\lambda (\frac {\hat {\psi}_0(p)}{p^2-(|k|+i0)^2},\hat {\psi}_0(p))=
\lambda \int _0^{2\pi}\int _0^{\pi}\int _0^{\infty}\frac {r^2{\rm sin}\theta |\hat {\psi}_0
(r,\phi ,\theta )|^2}{r^2-(|k|+i0)^2}d\phi d\theta dr=
$$
$$
=\frac {1}{2}\int _{{\bf R}}\frac {\xi (q)dq}{q^2-(|k|+i0)^2}=
\frac {1}{2|k|}\int _{{\bf R}}\frac {\xi (q)dq}{q-|k|-i0}=
$$
$$
=\frac {1}{2|k|}(i\pi \xi (|k|)+\int _{{\bf R}}\frac {\xi (q)dq}{q-|k|})=
\frac {i\pi}{2|k|}(1+S)(\xi )(|k|).
$$
By (8) and (11), we obtain (10). The lemma is proved.
\vskip 0.5cm

We note that the function $F$ can be extended to the whole real axis by means of the right-hand
side of the relation (10). Further, if the denominator of the expression (10) is equal 
to zero, then the numerator of this expression is equal to zero too. Therefore,
for all $q\in {\bf R}$ the relation (10) can be rewritten in the form
$$
\pi (iqF(q)+4\pi )\xi (q)+i\pi qF(q)(S)(\xi )(q)=-2q^2F(q). \eqno (12)
$$
We introduce the notation
$$
a(q)=2\pi (iqF(q)+2\pi ), \, b(q)=4\pi^2.
$$
Then (12) reduces to the equation
$$
(\frac {a(q)}{2}(1+S)+\frac {b(q)}{2}(1-S))\xi (q)=-2q^2F(q).  \eqno (13)
$$
\section {Uniqueness of the solution of the inverse problem}
\vskip 0.5cm

{\sc Lemma 2}.{\it The following relation
$$
(R_0(|k|)\psi _0,\psi _0)\to 0, \, |k| \to \infty 
$$
holds.}
\vskip 0.5cm

{\sc Proof}. We have the relation
$$
(R_0(|k|)\psi _0,\psi _0)=\frac {1}{4\pi}\int _{{\bf R}^3}(\int _{{\bf R}^3}
\frac {e^{i|k||x-y|}}{|x-y|}\psi _0(y)dy)\overline {\psi _0 (x)}dx=
$$
$$
=\frac {1}{4\pi}\int _{{\bf R}^3}(\int _{{\bf R}^3}
\frac {e^{i|k||y|}}{|y|}\psi _0(x-y)dy)\overline {\psi _0 (x)}dx=
$$
$$
=\frac {1}{4\pi}\int _{{\bf R}^3}
\frac {e^{i|k||y|}}{|y|}(\psi _0(-y)*\overline {\psi _0 (y)})dy. \eqno (14)
$$
It follows from the inclusion $\psi _0\in L^1({\bf R}^3)\cap L^2({\bf R}^3) $
that
$$
\psi _0(-y)*\overline {\psi _0 (y)}\in L^1({\bf R}^3)\cap L^{\infty}({\bf R}^3).
$$
Therefore,
$$
\frac {1}{|y|}\psi _0(-y)*\overline {\psi _0 (y)}\in L^1({\bf R}^3).  \eqno (15)
$$
By virtue of (14),
$$
(R_0(|k|)\psi _0,\psi _0)=\frac {1}{4\pi}\int _0^{\infty}e^{i|k|\rho}f(\rho )d\rho 
                                                     \eqno (16)
$$
where
$$
f(\rho )=\rho \int _0^{2\pi}d\psi \int _0^{\pi}d\theta {\rm sin}\theta (\psi _0(-(\cdot ))* 
\overline {\psi _0(\cdot )})(\rho ,\phi ,\theta )\in L^1(0,\infty ).
$$
(The last inclusion follows from the relation (15) in the spherical coordinates).
As well-known, the Fourier transform of the function beloning to $L^1({\bf R}^3)$
tends to zero when its argument tends to infinity. In view of (16), lemma is proved.
\vskip 0.5cm

{\sc Lemma 3}. {\it The inclusion $q^2F(q)\in L^2({\bf R})$ holds. In addition,}
$$
q^2F(q)\to 0, \, |q|\to \infty .
$$
\vskip 0.5cm

{\sc Proof}. Because of (8),(9), we have
$$
q^2F(q)=-\frac {2\pi ^2\xi (q)}{1+\lambda (R_0(q)\psi _0,\psi _0)}. \eqno (17)
$$
Therefore, to prove the first part of the lemma, as follows from 
Lemma 2, it is sufficient to establish
 that $\xi \in L^2({\bf R})$. Applying the Schwarz inequality
to the relation (9), we obtain
$$
\int _{\bf R}|\xi (q)|^2dq\leq C(\int _0^{2\pi}d\phi \int _0^{\pi}d\theta {\rm sin}
\theta \int _0^{\infty}q^4|\hat {\psi}_0(q,\phi ,\theta )|^4dq)^{1/2}=
$$
$$
=C(\int _{{\bf R}^3}|p|^2|\hat {\psi}_0(p)|^4dp)^{1/2}\leq C(\int _{{\bf R}^3}|p|^4
|\hat {\psi}_0(p)|^4dp)^{1/4}(\int _{{\bf R}^3}|\hat {\psi}_0(p)|^4dp)^{1/4}. \eqno (18)  
$$
By (6), we have $|p|\cdot |\hat {\psi}_0(p)|\in L^4({\bf R}^3)$. From here in view 
of the relation (5), it follows that the expression obtained in (18) is finite.
This implies the inclusion $\xi \in L^2({\bf R})$.

Now, we prove the second part of the lemma. According to the third condition in (4),						      
$$
|\widehat {\frac {\partial \psi _0}{\partial x_j}}(0)|=|p_j\hat {\psi}_0(p)|\to 0,\, 
|p|\to \infty .
$$
Thus, $|p|\cdot |\hat {\psi}_0(p)|\to 0$, as $|p|\to \infty$. By virtue of 
(9), we obtain
that $\xi (q)\to 0$, as $|q|\to \infty$. Because of Lemma 2, this proves the desired result. 
The lemma is proved.
\vskip 0.5cm

We mean by an index ${\rm ind}_{{\bf R}}\, a(q)$ of a complex-valued function $a(q)$ 
the number of
anticlockwise revolutions of the vector $a(q)$ in the complex plane when $q\in {\bf R}$
varies from $-\infty$ to $\infty$.
\vskip 0.5cm

{\sc Theorem 1}. {\it Assume that for the given scattering amplitude f the following
relations
$$
iqF(q)+2\pi \neq 0,\, q\in {\bf R},
$$
$$
{\rm ind}_{{\bf R}}\,(iqF(q)+2\pi )\geq 0
$$
holds. Then, the scattering amplitude $f$ corresponds not more than one potential in the
class of the potentials of the form $V=\lambda (\cdot ,\psi _0)\psi _0$ 
where $\psi _0$ satisfies (4).}   
\vskip 0.5cm
 
{\sc Proof}. First, we prove that the function $qF(q)$ is continuous. Note, that  the function 
$\hat {\psi}_0$ and, hence, the function 
$\xi$ has the continuous derivative. Further,
the operator $S$ transform the H\"older functions into the ones. Consequently, the function 
$(S\xi )(q)$ is continous. It follows from here by virtue of the relations (17),(11), that the 
function $qF(q)$ is continuous for $q\neq 0$. The continuity of the function $qF(q)$ at zero is 
the consequence of the relations (17),(14) and the last condition in (4). Therefore, 
the function $a(q)$ in the Eq.(13) is continuous. In addition, by Lemma 3, the limits of 
the function $a(q)$ for $q\to \pm \infty$ are the same and the right-hand side of Eq.(13) 
belongs to $L^2({\bf R})$.
Thus, we can apply to this equation, taking into account the condition of the theorem,
Th.6.1 in [2] for the space $L^2({\bf R})$. It follows from this theorem, 
that the number of the solutions of Eq.(13) in the class $L^2({\bf R})$ is not more than unity.
Assume that the functions $\pm \sqrt {|\lambda _j|}\psi _{0j}$ where $\pm$ is the sign of 
$\lambda _j$ generate the potentials $V_j,\, j=1,2$, corresponding to the 
unique amplitude $f$. What has been said above, implies, in view of (9),(11), that 
${\rm sign}\lambda _1={\rm sign}\lambda _2$ and the denominators on the right-hand side of (8)
for $\psi _0=\psi _{0j}, \, j=1,2$, are equel. Therefore, the correspondig numerators are
equel. Setting in (8) $n=\omega$, we obtain that
$|\sqrt {\lambda }\hat {\psi}_{01}(q)|=|\sqrt {\lambda }\hat {\psi}_{02}(q)|$. Let us write
the quantities $\sqrt {|\lambda |}\hat {\psi}_{0j}(q)$ in the trigonometric form:
$$\sqrt {|\lambda |}\hat {\psi}_{0j}(q)=|\sqrt {\lambda }\hat {\psi}_{0j}(q)|e^{i\theta _j(q)},
$$
where $\theta _j(q)$ are the arguments of $\sqrt {|\lambda |}\hat {\psi}_{0j}(q)$.
Then, in view of the above and (8), 
$$
e^{i(\theta _1(q)-\theta _1(q_0))}=e^{i(\theta _2(q)-\theta _2(q_0))},
$$
where $q=|k|n, q_0=|k|\omega$. From this for the fixed $q_0$, we obtain
$$\sqrt {|\lambda |}\hat {\psi}_{01}(q)=\sqrt {|\lambda |}\hat {\psi}_{02}(q)e^{i\theta _0},$$
where $\theta _0=\theta _1(q_0)-\theta _2(q_0)$. Therefore, the functions $\hat {\psi}_{0j}, j=1,2$
determinate not more than one potential. The theorem is proved.
\vskip 0.5cm

{\sc Corollary}. {\it Let $|qF(q)|<2\pi , q\in {\bf R}$, then Th.1 is true.}
\vskip 0.5cm

{\sc Remark}. The determination of the function $\psi _0$ via the scattering amplitude actually
means the calculation of the solution of the singular integral equation (13). In the case
$\kappa <0$ the dimension of the space of the solutions of the corresponding homogeneous 
equation coinsides with $|\kappa |$ [2]. Thus, in this case the Eq.(13) is not solved uniquely.
\vskip 0.5cm

We note, that by virtue of Lemma 3, there exists $A>0$ such that $|q|\geq A$ imlies
$$
|\frac {iqF(q)}{iqF(q)+4\pi}|\cdot \Vert S\Vert_{L^2({\bf R})}<1. \eqno (19)
$$
\vskip 0.5cm

{\sc Theorem 2}. {\it To any scattering amplitude $f$ there corresponds not more than one
potential in the class of potentials of the form $V=\lambda (\cdot ,\psi _0)\psi _o$ 
where $\psi _0$ satisfies (4) and also the condition}
$$
{\rm supp}\, \hat {\psi}_0\subset [A,\infty ).
$$
\vskip 0.5cm
{\sc Proof}. We rewrite Eq.(12) in the form
$$
\xi (q)+\frac {iqF(q)}{iqF(q)+4\pi}(S\xi )(q)=-\frac {2q^2F(q)}{\pi (iqF(q)+4\pi)}.
$$
We regard this equation in the class $L^2[A,\infty )$ changing $S\xi$ by $(S\xi )|_{[A,
\infty )}$. By (19), this equation always has a unique solution.
The completion of the proof can be derived similarly to Th.1.

\markright{}
\vskip 0.5cm
{\sc Список литературы}

{\em [1] \/} Newton R.G.: Scattering Theory of Waves and Particles.
New York: McGraw-Hill Book Company, 1966.

{\em [2] \/} Gohberg I.Ts., Krupnik N.Ya.: Introduction to Theory of 
One-Dimensional Singular Integral Operators. Kishin\"ov: Publishing Office
"Shtinitsa", 1973 (in Russian).

{\em [3] \/}  Reed, M., Simon, B.:
Methods of Modern Mathematical Physics. II.
Fourier Analysis, Self-Adjointness. New York: Academic Press, 1975.

\newpage
     Chuburin Yurii Pavlovich\\

     Physical-Technical Institute, Russian Academy of Sciences,\\

Ural Branch, Kirov Street 132, Izhevsk, 426001, Russia\\

     Phone: (3412) 21-89-88\\

     E-male: chuburin@otf.pti.udm.ru\\

%\ \ \ \ \ \ \ Summary\\
%\begin{center}
%INVERSE SCATTERING PROBLEM FOR THE SCHR\" ODINGER
%OPERATOR WITH A SEPARABLE POTENTIAL 
%\\

%Yu.P.Chuburin\\
%\end{center}
\newpage 
\ \ \ \ \ \ \ KEY WORDS

Schr\"odinger Operator

separable potential

inverse scattering problem

singular integral equation

\end{document}